%% file: main.tex
\documentclass[sigchi]{acmart}
\AtBeginDocument{%
  \providecommand\BibTeX{{%
    \normalfont B\kern-0.5em{\scshape i\kern-0.25em b}\kern-0.8em\TeX}}}

\setcopyright{acmcopyright}
\copyrightyear{2022}
\acmYear{2022}
\acmDOI{10.1145/3490099.3511146}

\copyrightyear{2022}
\acmYear{2022}
\setcopyright{rightsretained}
\acmConference[IUI '22]{27th International Conference on Intelligent User Interfaces}{March 22--25, 2022}{Helsinki, Finland}
\acmBooktitle{27th International Conference on Intelligent User Interfaces (IUI '22), March 22--25, 2022, Helsinki, Finland}\acmDOI{10.1145/3490099.3511146}
\acmISBN{978-1-4503-9144-3/22/03}

\usepackage{amsmath}
\usepackage{wrapfig}
\usepackage{bbold}
\newcommand{\app}{iSEA}

\begin{document}

\title{\app: An Interactive Pipeline for Semantic Error Analysis of NLP Models}

\author{Jun Yuan}
\email{junyuan@nyu.edu}
\affiliation{
\institution{New York University}
\country{USA}
}

\author{Jesse Vig}
\email{jvig@salesforce.com}
\affiliation{
\institution{Salesforce Research}
\country{USA}
}

\author{Nazneen Rajani}
\email{nazneen.rajani@salesforce.com}
\affiliation{
\institution{Salesforce Research}
\country{USA}
}



\renewcommand{\shortauthors}{Yuan et al.}

\begin{abstract}
    \input{sections/00-abstract}
\end{abstract}

\ccsdesc[500]{Human-centered computing~Interactive systems and tools}
\ccsdesc[500]{Computing methodologies~Natural language processing}
\keywords{Error Analysis, xAI}



\maketitle

\input{sections/01-intro}
\input{sections/02-related}
\input{sections/03-pipeline}

\input{sections/04-error}

\input{sections/05-system}
\input{sections/06-case}

\input{sections/07-limit+conclude}

\bibliographystyle{ACM-Reference-Format}
\bibliography{ref}

\end{document}

%% file: sections/00-abstract.tex
Error analysis in NLP models is essential to successful model development and deployment. One common approach for diagnosing errors is to identify subpopulations in the dataset where the model produces the most errors. However, existing approaches typically define subpopulations based on pre-defined features, which requires users to form hypotheses of errors in advance. To complement these approaches, we propose {\app}, an Interactive Pipeline for Semantic Error Analysis in NLP Models, which automatically discovers semantically-grounded subpopulations with high error rates in the context of a human-in-the-loop interactive system. {\app} enables model developers to \textit{learn} more about their model errors through discovered subpopulations, \textit{validate} the sources of errors through interactive analysis on the discovered subpopulations, and \textit{test hypotheses} about model errors by defining custom subpopulations. The tool supports semantic descriptions of error-prone subpopulations at the token and concept level, as well as pre-defined higher-level features. Through use cases and expert interviews, we demonstrate how {\app} can assist error understanding and analysis.


%% file: sections/01-intro.tex
\section{Introduction}

Being able to understand and analyze when and why a model makes mistakes is essential for developing accurate and robust NLP models. One common approach for error analysis is to identify the \textit{subpopulations}, or subsets, of the dataset where the error rate is high. Typically these subpopulations are constructed based on higher-level pre-defined features, for example document length~\cite{li2018textbugger} (e.g. the subpopulation of examples with length less than 10 words), or part-of-speech tags ~\cite{joshi2018shot} (e.g. the subpopulation contain the second person pronoun).
While such pre-defined criteria may be effective at identifying certain classes of errors, they are not able to capture the full range of error conditions, in particular those errors that are grounded in specific semantic concepts. Moreover, they require someone with prior knowledge of a domain to form hypotheses about error causes in order to construct such features.

Some recently developed tools, such as Errudite~\cite{wu2019errudite}, enable users to define a rich array of custom rules for extracting subpopulations, including word-level features for capturing semantically meaningful subpopulations. However, users must learn a new query language to define such subpopulations and must have sufficient prior knowledge on the model to form relevant queries. Other interactive tools, such as LIT~\cite{tenney2020language}, enable users to select an instance of interest and then derive a group of similar instances for analysis. However, there is no interpretable description of such a group and thus users need to manually inspect the group to determine its characteristics.

To address the limitations of existing error analysis tools for NLP models, we introduce {\app}\footnote{\url{https://github.com/salesforce/iSEA}}, an \textbf{I}nteractive Pipeline for \textbf{S}emantic \textbf{E}rror \textbf{A}nalysis of NLP Models. Rather than constructing subpopulations based on a narrow set of predefined features, {\app} defines subpopulations based on individual words present in the input text (e.g. all inputs containing ``delicious''), or a collection of words defining a semantic concept (e.g. all inputs containing ``delicious'', ``tasty'', or ``yummy''). Most importantly, the tool supports automated discovery of the subpopulations with the highest error rates, so that no advanced knowledge of the task or domain is required. To support human insight and control over the process, the tool embeds this automated discovery process within a human-in-the-loop-pipeline: users may modify and iterate on the discovered subpopulations and also define custom subpopulations. The pipeline is instantiated in a graphical user interface that also supports exploration of the data and interactive analysis. We evaluate {\app} based on interviews with domain experts and exploration of two use cases.

In summary, our contribution in this work includes: (1) an human-in-the-loop pipeline for semantic error analysis of NLP models, (2) an interactive visual analytics system as an instantiation of this pipeline, and (3) use cases on two NLP tasks and domain expert interview as an evaluation of {\app}.


The paper is organized as follows. Section 2 presents background and related work. Section 3 provides an overview of the pipeline and the design goals. Section 4 describes in detail the methods for error discovery and validation, as well as the system architecture of {\app}. Section 5 introduces the {\app} interface. Section 6 evaluates the usefulness of the system by two hypothetical usage scenarios and interviews with three domain experts. The societal impact of this work and limitations are discussed in Section 7 and Section 8. 

%% file: sections/02-related.tex
\section{Background and Related Work}


In recent years, there has been increasing interest in interactive tools that help users understand where their models are failing. Being able to understand errors in a model is important for robustness testing~\cite{goel2021robustness}, improving overall performance ~\cite{wu2019errudite}, and increasing user trust~\cite{arendt2020crosscheck}. 
Many methods and systems are introduced for model debugging and model diagnosis ~\cite{alsallakh2014visual,bilal2017convolutional,goel2021robustness,kahng2017cti,tenney2020language,wang2019deepvid,zhang2018manifold}. However, these works support error analysis by enabling users to \textit{select} or \textit{filter} instances by pre-defined metrics, and then \textit{understand} the model behavior, in order to \textit{diagnose} the model weakness. Our work, instead, tries to automatically discover the error-prone subpopulations so that the users can first \textit{learn} where the errors happen, then \textit{validate} the potential error cause proposed by the system and further \textit{test} their own hypothesis on error causes. In the latter part of this section, we discuss existing error analysis approaches for NLP models and methods that suggest error-prone subpopulations.

\textbf{Subpopulation-level Error Analysis for NLP Models: } Inspecting model errors from a perspective of subpopulation is common for NLP tasks. For example, in QA tasks, researchers have analyzed the error distribution across question types ~\cite{liu2017stochastic} and document length~\cite{li2018textbugger}. In recent years, research has increasingly focused on subpopulation-level error analysis. \textit{Errudite}~\cite{wu2019errudite}, for example, introduces a custom query language to explore subpopulations of interest. However, constructing these queries requires users to form hypotheses of potential causes of errors in advance. In our work, we instead provide users with initial ``hypotheses'' by automatically extracting error-prone subpopulations across the broad range of lexically-defined groupings. \textit{Checklist}~\cite{ribeiro2020beyond} also provides suggestions for which subpopulation to inspect. However, the goal of \textit{Checklist} is assisting users to create testing sets to find potential bugs in a model and test the model robustness, while the goal of our work is to help users understand and analyze existing errors. LIT ~\cite{tenney2020language}, another interactive tool for explaining and inspecting NLP models, enables users to search and inspect a group of instances that are similar to a specific instance, but does not summarize \textit{how} these features are similar. 
Another recent work from Gururangan \textit{et. al} ~\cite{gururangan2018annotation} found that models may overfit to annotation artifacts in the training set and perform poorly on more realistic datasets. In our work, we try to assist users to understand what annotation artifacts the model has learned that may lead to more errors. \textit{Manifold}~\cite{zhang2018manifold}, also provides subpopulation-level model inspection for model understanding and diagnosis. However, it focuses more on model comparison and analysis of where models agree or disagree. Similarly, \textit{Responsible-AI-Widgets}~\cite{respobibleai} supports error analysis in subpopulations, but it is a general dashboard not customized for NLP data. In Section ~\ref{sec:method}, we introduced multiple features and principles of discovering and describing error-prone subpopulations in NLP data. 



\textbf{Automatic Error Discovery:}
Innovations in error analysis methods have come from both the machine learning and HCI communities. 

A recent work, \textit{FairVis}~\cite{cabrera2019fairvis} integrates a technique to automatically generate subpopulations with high error rates based on clustering. However, their work focus more on the notions of fairness and is designed for tabular data. Although \textit{FairVis} extracts ``dominant features'' to describe a discovered subpopulation, such description is just an approximation of the discovered subpopulation, which brings uncertainty to the understanding of a subpopulation. Chung \textit{et. al} ~\cite{chung2019slice} also pointed out that clustering usually finds arbitrary subsets for errors.

Different from the bottom-up approaches, which discover subpopulations and then summarize their characteristics, a top-down approach is to keep adding feature values as constraints of a subpopulation. For example, rule-based models which end up with a set of if-then rules can provide interpretable descriptions of different subpopulations. In recent years, rules have been widely used for text classification based on high-level lexical features~\cite{hutto2014vader}, syntactic- and meta-level ~\cite{chen2019clinical} features. However, automatic rule generation for error analysis is still remained to be explored. In our work, we apply token-level features in rules to provide semantic context for error analysis. Most similar to our work, Slice Finder~\cite{chung2019slice} automatically generates interpretable data slices (subpopulations) containing errors based on decision tree and breadth search. It keeps adding features for more granular groups until the training loss is statistically significant.
Our error discovery method is different in several key ways. First, we aim to analyze NLP models so that we apply several optimization techniques for text data. Not only do we support multiple levels of features, we also pre-process text data by feature importance to reduce the amount for searching, which accelerates the process of error discovery. Second, we take a different set of metrics for subpopulation discovery, including thresholds of error rates, number of features, and support (data coverage), which ensure us to use as few features as possible to find large enough subpopulations with high error rates.



%% file: sections/03-pipeline.tex
\section{Error Analysis Pipeline}
\label{sec:pipeline}
In this section we outline the design goals for a human-in-the-loop error analysis pipeline and provide an overview of the resulting pipeline (Fig.\ref{fig:pipeline}).  


\subsection{Design Goals}
\label{sec:task}
The goal of this work is to assist model developers and other users in understanding the errors made by an NLP model through a human-in-the-loop pipeline. More precisely, our objective is to guide users to understand, given a model and its input and output, where the model makes mistakes and to form hypotheses about why the model makes mistakes. 


We decompose the error analysis problem into the following design goals, which we express as a set of questions:

\textit{G1: What is the performance of the model?} 
The overall information of the model, data and error is necessary in order to place the performance of a specific subpopulation in context and gain an general understanding of error distribution.



\textit{G2: What aspects of the data does the model fail to understand?} 
For example, does the model fail to understand particular words, or a particular style of text? Being able to answer such questions may help the user develop actionable insights to improve the model.

\textit{G3: What characteristics of the training data might not generalize to an out-of-distribution dataset?}  Another cause of error comes from learning spurious correlation between a feature and a specific outcome in the training data that may not be present in the test set. Thus the tool should enable users to compare the model performance on in-distribution (ID) data and out-of-distribution (OOD) data.

\textit{G4: What is the model performance for a user-defined subpopulation?} Being able to test model performance on customized subpopulations enables users to test their hypotheses about the causes of errors and to investigate types of errors that have been observed in other models. For example, users may wish to better understand whether the model has biases related to gender or race~\cite{garrido2021survey,tan2019assessing,vig2020causal}.

\begin{figure*}
    \centering
    \includegraphics[width=\textwidth]{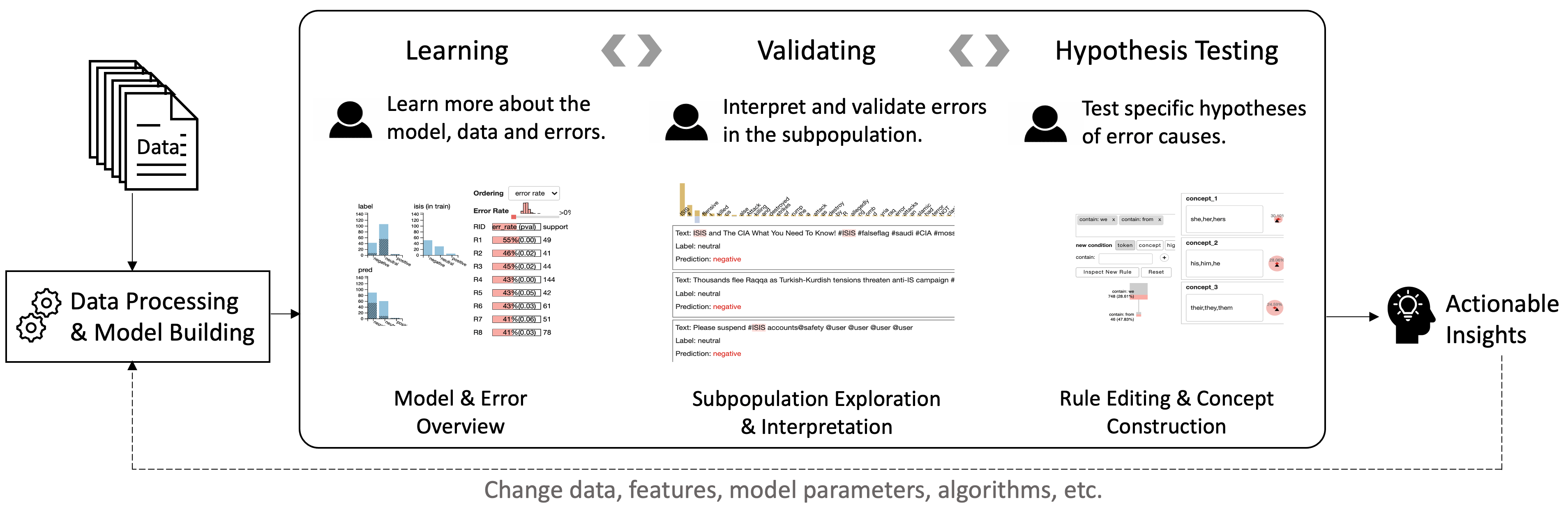}
    \caption{An overview of the human-in-the-loop pipeline for semantic error analysis.}
    \Description[the pipeline for semantic error analysis]{the human-in-the-loop pipeline for semantic error analysis include learning, validating, and hypothesis testing}
    \label{fig:pipeline}
\end{figure*}

The interactive pipeline has three main stages, each supported by visual interfaces as shown in Fig.~\ref{fig:pipeline}. 

\textbf{Learning.} The first stage focuses on discovery of error-prone subpopulations, as well as assessing overall model performance (\textbf{G1}). We  compute and present the descriptions of discovered subpopulations where the error rate is higher than the baseline error rate. We present the model performance disaggregated over several high-level features, for example document length and class label, using a set of bar charts.


\textbf{Validating.} The second stage is to further analyze specific subpopulations where the model makes more errors. The tool provides explanations that highlight the role of specific tokens within a subpopulation based on aggregated SHAP values (see Section~\ref{sec:method}). This can help users understand whether a particular word is contributing to the errors, or simply correlates with another concept that may be causing the errors (\textbf{G2}). Users may also manually inspect examples within the subpopulations to form their own opinions of the causes of the errors. The tool also supports comparison between the training and subpopulation distributions to help determine whether the errors are caused by OOD data (\textbf{G3}).

\textbf{Hypothesis Testing.} The third stage enables the users to test the model performance over a custom subpopulation. Users may define rules at different levels of granularity including token-level, concept-level, and metric-level, allowing them to easily test a specific hypothesis (\textbf{G4}).

Although error analysis usually starts from the \textit{learning} stage where users gain a general understanding of model performance and error distribution, users may enter the pipeline at any stage and finish their tasks in a flexible manner. For example, if a model developer is already familiar with the data and model behaviors, they may prefer to \textit{test hypotheses} directly and then \textit{validate} the generated insights.

To instantiate this pipeline, we developed {\app}, an interactive visual analytics tool for semantic error analysis in NLP models. The system supports the introduced human-in-the-loop pipeline and integrates all the features to reach the design goals, which we will describe in the following sections.

%% file: sections/04-error.tex
\section{Error Analysis Method}
\label{sec:method}

In this section, we describe the techniques used in {\app} to conduct semantic error analysis on NLP models, which consists primarily of two parts: (1) automated error discovery: identifying subpopulations where the model consistently makes mistakes, and (2) interpretation of errors: correlating the erroneous subpopulation with the model behavior explanations. We first describe the features we use to define subpopulations and the principles that guide us to extract and present rules. We then go through the methods of automatic error discovery and error interpretation. Finally, we introduce the overall system architecture that we employ to support interactive semantic error analysis.

\subsection{Features and Rule Presentation Principles}

The system of {\app} followed an iterative design process. In each iteration, we tested whether the extracted rules and the presented information can sufficiently answer the questions posed in Section~\ref{sec:task}. To better support the error analysis, we defined three types of features to describe subpopulations and four principles for more interpretable rule representation.

The three types of features for our pipeline are: \textit{token}, \textit{concept}, and \textit{high-level features}.
\textit{Tokens} refer to unigram (e.g., ``want''), bigram (e.g., ``want to'') and trigram (e.g., ``want to be'') in this work. 
Token-level features are semantically meaningful and may describe a broad range of subpopulations; therefore these are the main features we use to define subpopulations. 

A \textit{concept} is a group of semantically similar \textit{tokens}. For example, tokens that belong to the same named entity (e.g., country names), tokens that refer to pronouns related to female (e.g. ``she'', ``her'', ``hers''). This feature is mainly used for hypothesis testing as introduced in Section~\ref{sec:pipeline}. 

\textit{High-level features} include a list of pre-defined high-level metrics that are commonly used in error analysis,
such as document length~\cite{li2018textbugger}, part-of-speech tags ~\cite{joshi2018shot}, and word overlap (for the NLI task)~\cite{mccoy2019right}. 
{\app} supports error analysis on high-level features across the three stages we defined in the pipeline.

We also identified four principles of presenting rules to achieve human interpretability and ensure that the rules describe subpopulations with a significantly higher error rate.


\textbf{Limit the number of conditions.}
Subpopulations where several token-level features exist in the same document are usually small, which makes the description of errors unreliable. Long rules with too many features are also hard to interpret. As such, we limit the maximum number of conditions in each rule to two in this work.

\textbf{Test significance.} To ensure that a high error rate in a subpopulation does not occur by chance, we apply statistical significance testing. We provide two ways of testing significance: (1) calculating the p-value of the null hypothesis that the error rate in the subpopulation is higher than the baseline error rate, which is the error rate on the whole data set; (2) running bootstrap~\cite{diciccio1996bootstrap} to compute 95\% confidence interval of the error rate on a subpopulation. 

\textbf{Limit the cardinality of features.} Given a condition of ``\texttt{document length} > 23'', users need to understand whether such document length is actually long or short relative to other examples in the dataset. 
In this work, we use the value of $10th$ percentile and $90th$ percentile as the thresholds for high-level features to differentiate the three ranges, which makes \textit{low} and \textit{high} extreme cases that are important to error analysis.
As for the token-level features or concept-level features, we use binary values of whether such tokens or concepts appear in the document in order to provide an easily interpretable description of the subpopulation.

\textbf{Avoid negation.} One might define subpopulations based on the absence (negative value) of a particular feature, e.g. all documents that do \textit{not} contain ``\textit{blue}''. However, these types of rules are difficult to link to an interpretable concept or actionable insight on improving the model and are therefore not used in the tool.

\subsection{Error Discovery}
We view the problem of finding error-prone subpopulation as a classification task. Assume we have a model $M$ that generates predictions $\mathcal{Y}_{pred}=\{y_{pred}^{(i)}\}^N_{i=1}$ on a set of $N$ input document $\mathcal{X}=\{x^{(i)}\}^N_{i=1}$ with ground truth defined as $\mathcal{Y}_{gt}=\{y_{gt}^{(i)}\}^N_{i=1}$. To analyze the errors made by $M$, we define the labels of whether the documents are predicted incorrectly by $M$ as $\mathcal{Y}_{error}$:
\begin{equation}
    \mathcal{Y}_{error} = \{ [y_{pred}^{(i)} \neq y_{gt}^{(i)}]\}^N_{i=1}
\end{equation}

where $[y_{pred}^{(i)} \neq y_{gt}^{(i)}]$ evaluates to 1 if $y_{pred}^{(i)} \neq y_{gt}^{(i)}$ and 0 otherwise. By applying rule-based models, the extracted \textit{if-then} rules used to classifying whether a document is predicted incorrectly by $M$ becomes a set of descriptions for error-prone subpopulations.

However, applying existing rule-based models (or tree-based models) cannot fulfill the principles we introduced in the previous subsection. So in this work, we use a tree-based model, random forest, as a preliminary step of filtering important features, that is, features that are useful for describing an error-prone subpopulation. This is important for error discovery involving token-level features because of the large number of such features. After this step, we usually get fewer than $500$ tokens to process.

The algorithm of discovering error-prone subpopulations contains four steps. First, we train a random forest where we limit the max depth to 3 in order to accelerate the model training. Next, we filter the features with non-zero feature importance as candidate features for the next step. The two steps can be skipped for high-level features because the number of pre-defined high-level features is usually small. We then directly test the error rate in a subpopulation described by one feature (contains a token or not; or low/medium/high value of a high-level feature), as well as the combination of two or three features. Given a pre-defined minimal error rate and support threshold, we report the results with high error rate. Finally we test the significance of the difference between the discovered error-prone subpopulation and the full test set by computing p-values for the null hypothesis and the 95\% confidence intervals of the subpopulation error rate through boostrapping. In the implementation, we take the minimal error rate as the error rate over the entire test set, and the threshold for support as $5\%$ of the data.

\subsection{Model Behavior Explanation for Error Validation}
\label{sec:model_expl}
To better understand how the token or tokens that define a subpopulation actually contribute to the errors, we present explanations of model behaviors based on SHAP values~\cite{lundberg2017unified}, which indicate how strongly particular tokens influence the model predictions.
For a given class $\mathcal{C}$, e.g. \textit{positive sentiment}, a token with positive SHAP value means that this token increases the probability of predicting this document as class $\mathcal{C}$, and decreases the probability if the SHAP value is negative.
We use SHAP values instead of other methods such as LIME~\cite{ribeiro2016should} and DeepLift~\cite{shrikumar2016not} because SHAP has better computational performance and has constraints to ensure that they have better consistency with human intuition ~\cite{lundberg2017unified}, which are important in a human-in-the-loop pipeline. 

To give an overview of how the tokens are relevant to the model predictions, we also provide a subpopulation-level model explanation aggregating the SHAP values on all the documents in a given subpopulation. Given $SHAP(t|x, \mathcal{C})$ as the SHAP value of the token $t$ in the input $x$ for a given class $\mathcal{C}$, we define $cnt_{pos}(t)$ as the number of documents that contains token $t$ which has a positive SHAP value for model prediction of class $\mathcal{C}$:
\begin{equation}
    cnt_{pos}(t) = \sum_{x \in \mathcal{X}^*, t \in x} [SHAP(t|x, \mathcal{C})>0]
\end{equation}
where $\mathcal{X}^*$ is a given subpopulation, and $[SHAP(t|x, \mathcal{C})>0]=1$ if $SHAP(t|x, \mathcal{C})>0$, 0 otherwise. Similarly, we define $cnt_{neg}(t)$ as:
\begin{equation}
    cnt_{neg}(t) = \sum_{x \in \mathcal{X}^*, t \in x} [SHAP(t|x, \mathcal{C})<0]
\end{equation}

In the implementation, for each class we keep only the top three tokens in each document with the highest absolute SHAP values so that we can calculate and render such subpopulation-level model explanations in real time. 

\subsection{System Architecture}
\begin{figure}
    \centering
    \includegraphics[width=.5\textwidth]{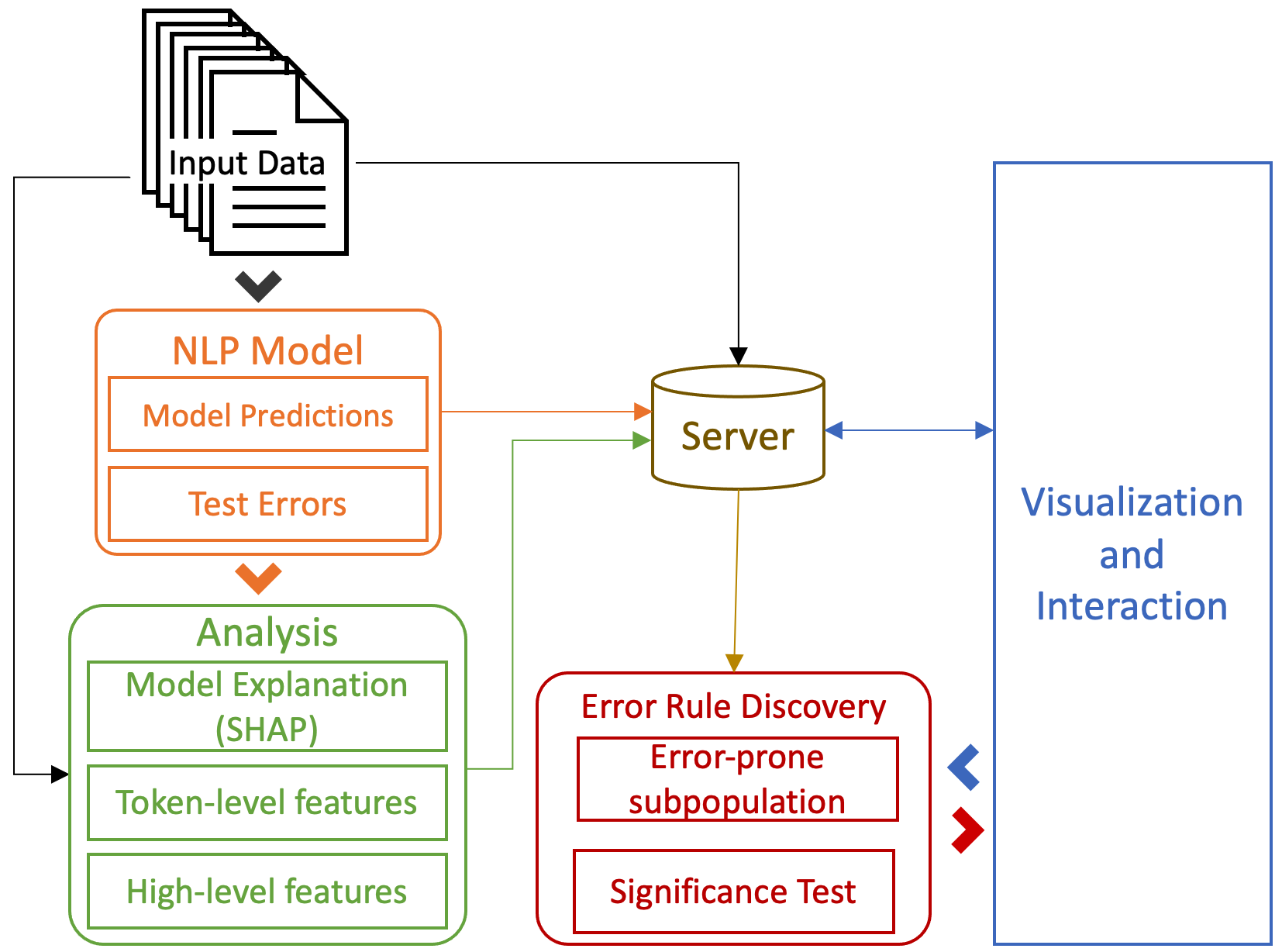}
    \caption{The system architecture of {\app}.}
    \Description[The system architecture of {\app}]{The system architecture of {\app}}
    \label{fig:data_process}
\end{figure}

Fig.~\ref{fig:data_process} illustrates the system architecture and the workflow of data processing for {\app}, which contains four main modules: (1) the NLP model module, (2) the analysis module, (3) the error rule discovery module, and (4) the visualization and interaction module.  

In particular, in the NLP model module, the input documents are passed into the given model, which outputs model predictions $\mathcal{Y}_{pred}$ that are used to compute model errors $\mathcal{Y}_{error}$. The analysis module extracts the token-level features and the pre-defined high-level features from the input documents, and then computes the SHAP values as model explanations for later aggregation and analysis. The error rule discovery module generates rules to describe error-prone subpopulations using the extracted features and the model error labels derived from the previous modules, and then reports the significance test results of the error rates in the discovered subpopulations. All pre-process data is cached on the server side. The visualization and interaction module employs several coordinated views to assist the interactive error analysis of the given NLP model. Various interactions are supported by the real-time communication between the user interface (visualization and interaction module) and the server. All the modules work together to form a human-in-the-loop error analysis mechanism to understand and diagnose the behavior of an NLP model.

%% file: sections/05-system.tex
\section{Visualization and User Interface}

\begin{figure*}
    \centering
    \includegraphics[width=\textwidth]{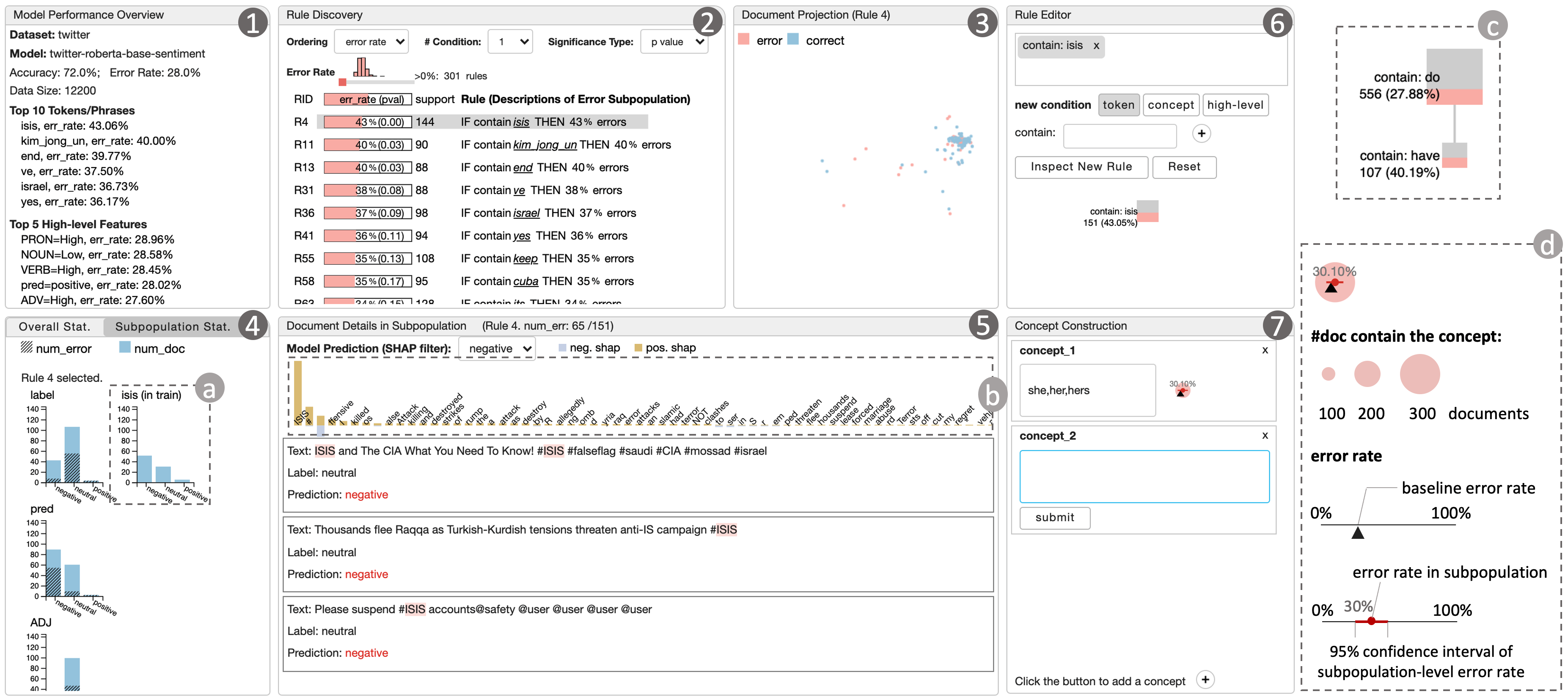}
    \caption{iSEA contains seven linked views: (1) Model Performance View, (2) Rule Discovery View, (3) Document Projection View, (4) Statistics View, (5) Document Detail View, (6) Rule Editing View, (7) Concept Construction View.}
    \Description[iSEA contains seven linked views]{iSEA contains seven linked views}
    \label{fig:ui}
\end{figure*}

\subsection{User Interface Overview}
As shown in Fig.~\ref{fig:ui}, the {\app} interface provides seven views to assist users to semantically understand and analyze errors in an NLP model: (1) the \textit{model performance view} provides an overview of model performance such as tokens and high-level features that are relevant to a high error rate; (2) the \textit{rule discovery view} presents a list of automatically extracted rules and the error rate in the described subpopulation; (3) the \textit{document projection view} shows the distribution of documents based on the techniques of sentence embedding~\cite{reimers2019sentence} and t-SNE~\cite{van2008visualizing} as well as whether the documents are predicted correctly or not by the model; (4) the \textit{statistics view} shows how the errors distribute across pre-defined high-level features on the whole data set (the tab of \textit{Overall stat.}) and on a specific subpopulation (the tab of \textit{Subpopulation stat.}); (5) the \textit{document detail view} displays the actual documents that match a selected/created rule and integrated SHAP values to assist users to understand how the model makes decisions in a subpopulation; (6) the \textit{rule editing view} allows users to edit the discovered rules or create their own rules using tokens, concepts or high-level features; (7) the \textit{concept construction view} enables the construction and comparison of user-defined concepts. All the views are interactively connected to assist error analysis through the three stages of (1) \textit{learning} the general information of the model and the data, as well as where errors may happen, (2) understanding and \textit{validating} the actual causes of errors, (3) \textit{testing hypothesis} of model performance based on user defined rules.
In the following subsections, we introduce the coordinated views according to how they support the three stages in the pipeline.


\subsection{Learning: Model, Data, and Errors}

The \textit{model performance view} (Fig.~\ref{fig:ui}\textcircled{1}) provides an overview of the model and data, including the overall accuracy, the baseline error rate, as well as a preview of tokens and high-level feature values that
describe subpopulations with a high error rate. By reading the information in this view, users can gain a general understanding of the model performance and the potential causes of errors.

In the \textit{rule discovery view} (Fig.~\ref{fig:ui}\textcircled{2}), we present the complete list of automatically extracted rules and the error rates of the respective subpopulations. For each rule, we also show a p-value that indicates the probability that an observed difference between the subpopulation-level error rate and the baseline error rate could have occurred by chance; the smaller the p-value, the greater the statistical significance of the observed difference. Alternatively, users can choose to view the $95\%$ confidence interval of the subpopulation-level rate; the smaller the range, the greater the statistical significance of the observed error rate. The \textit{support} of a rule---the size of the associated subpopulation---is also shown at the beginning of each rule to help users understand the frequency of the error pattern. 

To provide an overview of these automatically extracted rules, a histogram of the error rates is shown on the top of the view, which also provides a slider for filtering rules based on error rate. Additionally, {\app} supports filtering by number of conditions and ordering rules based on either rule support or error rate to assist users in finding subpopulations of interest.

\subsection{Error Understanding and Validating}
The \textit{document projection} together with views at the bottom (Fig.~\ref{fig:ui} \textcircled{3}\textcircled{4}\textcircled{5}) aim at helping users understand model behaviors and the document content to \textit{validate} the causes of errors (\textbf{G2, G3}).

The \textit{document projection view} (Fig.~\ref{fig:ui}\textcircled{3}) on the top provides an overview of the document distribution. In this view, each point represents a document in the data set, and the color indicates whether this document is predicted correctly by the model. To map the documents to a 2-dimensional space, we first extract document embedding vectors using Sentence Transformers~\cite{reimers2019sentence}, a state-of-the-art framework of computing dense vector representations for sentences, paragraphs, and images based on their semantic meaning.  Then we apply t-SNE~\cite{van2008visualizing}, a dimensionality reduction technique, to project the high-dimensional document embedding vectors to a 2-dimensional space. These techniques ensure that semantically similar documents are also closer in the 2D space.

\begin{figure}
    \centering
    \includegraphics[width=0.35\textwidth]{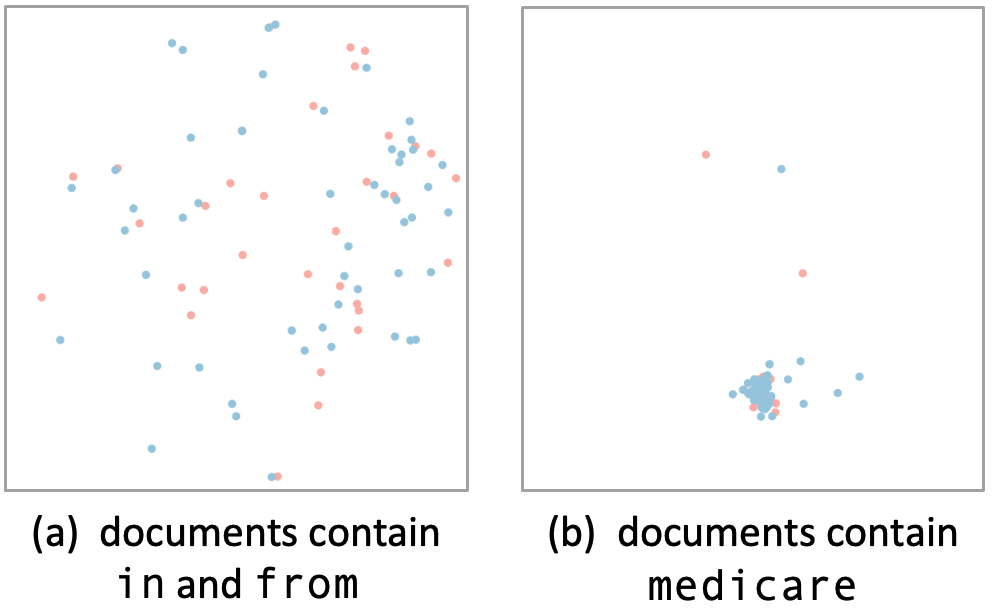}
    \vspace*{-.3cm}
    \caption{The projection of documents from two different subpopulations.}
    \Description[The two dimensional projection of documents from two different subpopulations.]{The documents containing ``in'' and ``from'' are distributed everywhere in the projection, but the documents containing ``medicare'' are clustered together in the projection.}
    \label{fig:projection}
    \vspace*{-.2cm}
\end{figure}

When there is no rule selected or created by the user for inspection, this view presents the distribution of the documents from the entire test set. Once the user selects or creates a rule for analysis, this view shows the distribution of documents in the corresponding subpopulation, enabling users to better understand the semantic relationships, as illustrated by the example in Fig.~\ref{fig:projection}. The documents in Fig.~\ref{fig:projection}(a) are more distributed, indicating that these documents are not semantically similar, while documents shown in Fig.~\ref{fig:projection}(b) tend to cluster together because all of them mention the word \texttt{medicare} and are similar in terms of their semantic meanings. 

At the bottom of the interface, the \textit{statistics view} (Fig.~\ref{fig:ui}\textcircled{4}) and \textit{document view} (Fig.~\ref{fig:ui}\textcircled{5}) support further validation of error causes through feature disaggregation, posthoc model explanations, and manual inspection of documents. 

In the \textit{statistics view}, we show the number of errors across labels, model predictions, as well as other high-level features. Under the tab of \textit{Overall stat.}, the statistics are based on the errors on the entire test set. Once the user selects or creates a specific rule, the statistics for that subpopulation will be shown under the tab of \textit{Subpopulation stat}. As shown in Fig.~\ref{fig:ui}a, we also provide the distribution of the tokens mentioned in a rule across the labels in the training set.

The \textit{document detail view} provides a bar chart of aggregated SHAP values for documents in a subpopulation (Fig.~\ref{fig:ui}b) and shows the actual documents below the chart. The integrated bar chart displays the model explanation as described in Section~\ref{sec:model_expl}. We use the height of bars in yellow to represent the value of $cnt_{pos}$, and the height of blue bars for $cnt_{neg}$ in an opposite direction at the same column. To highlight the tokens that are frequently considered as important by the model, we sort the columns (tokens) in a descending order by either the value of $cnt_{pos}$ or $cnt_{neg}$ according to whether there are more documents contain tokens with positive SHAP values or negative ones. 

As for the list of the documents, we show the incorrectly predicted documents in the beginning of the list. In each document, we highlight in red the tokens used in the rule. These features help users to quickly find the documents on which the model makes mistakes and focus on the potential error causes mentioned in a rule.

\subsection{Hypothesis Testing}
Once the users gain enough knowledge about the model and the data, they can create rules to test their own hypotheses (\textbf{G4}) through the views on the right-hand side (Fig.~\ref{fig:ui}\textcircled{6}\textcircled{7}), and then further validate them through the views described in the previous subsection. 

The \textit{rule editing view} (Fig.~\ref{fig:ui}\textcircled{6}) allows the users to edit any selected rules or create a new rule. Any selected rule will be added to this view directly so that users can add or remove conditions based on extracted rules.
Users may also create their own rules by clicking ``Reset'' to clear all the existing conditions. {\app} supports the combination of three types of features: token, concept, and high-level features.

The \textit{concept construction} view (Fig.~\ref{fig:ui}\textcircled{7}) enables users to add a new concept or edit an existing concept. Initially, a newly added concept box is editable as shown in Fig.~\ref{fig:ui}\textcircled{7}:concept\_2. A submitted concept (Fig.~\ref{fig:ui}\textcircled{7}:concept\_1) shows the user-defined token/phrase list for this concept, as well as a summary visualization of the subpopulation that contains the concept. The summary visualization is illustrated in Fig.~\ref{fig:ui}c. Such visualization enables visual comparison of subpopulations containing different concepts to test the hypothesis related to multiple concepts. 

%% file: sections/06-case.tex
\section{Evaluation}
To demonstrate how {\app} can help with ML model development and deployment, we present usage scenarios on two NLP tasks: natural language inference (NLI) and sentiment analysis. The design goals set out in Section ~\ref{sec:task} are used to evaluate the performance of the tool and are referenced directly in the use cases. We then present interviews with three domain experts in order to illustrate how {\app}  is used in practice.

\begin{figure*}
    \centering
    \includegraphics[width=\textwidth]{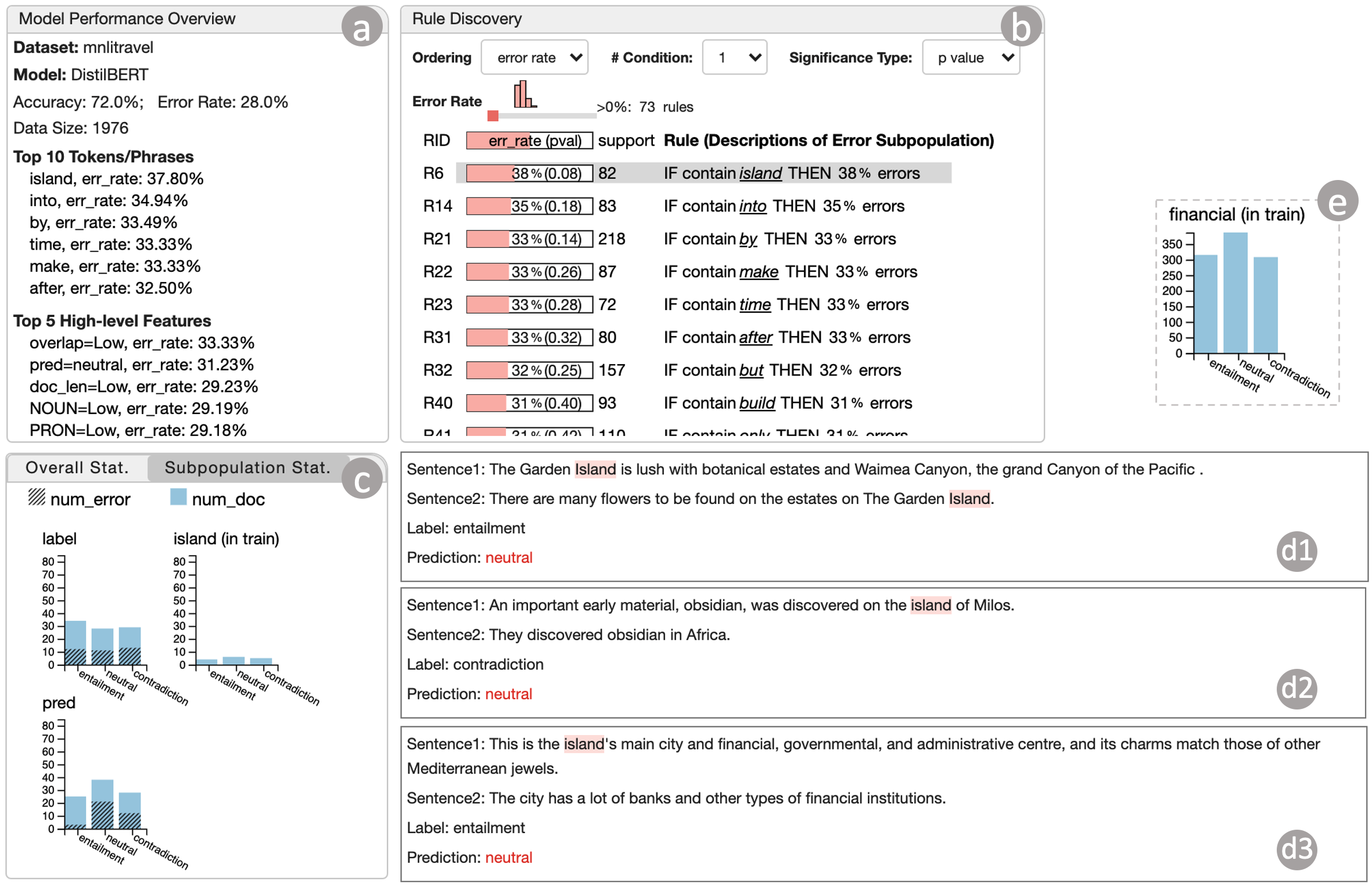}
    \caption{Usage scenario: Testing the robustness of a DistilBERT model for NLI task on an OOD dataset of travel genre.}
    \Description[The user interface for usage scenario 1]{The user interface of analyzing errors from a DistilBERT model for MNLI dataset}
    \label{fig:mnli}
\end{figure*}

\subsection{Usage Scenario: Natural Language Inference (NLI)}
\label{sec:nli}
In this hypothetical scenario, we show how {\app} helps model developers understand the robustness of their model by analyzing the model errors on an out-of-distribution (OOD) dataset. 

A model developer, Alice, trains a DistilBERT model~\cite{sanh2019distilbert} on a set of documents from the \textit{government} genre in the Multi-NLI dataset~\cite{N18-1101}. This model will be used for an NLI task: given a \textit{premise} sentence and a \textit{hypothesis} sentence, the model predicts whether the \textit{hypothesis} is an \texttt{entailment}, or a \texttt{contradiction} of the \textit{premise}, or just \texttt{neutral}. At present, this model can reach $80\%$ accuracy on the test set. Alice wants to know the robustness of this model so she tests the model on a data set from the \textit{travel} genre in Multi-NLI dataset. This test ends up with only $72\%$ accuracy, which shows a drop of $8\%$ for this OOD data compared to the in-distribution data.

To understand the accuracy drop and analyze the error causes, Alice first gains a general understanding of where the model makes mistakes (\textbf{G1}) from the \textit{model performance view} as shown in Fig.~\ref{fig:mnli}a,  which provides an overview of the data set and potential causes of errors. Under \textit{Top 10 Tokens/Phrases}, which shows the tokens that define subpopulations with the highest error rates, she sees that the first word is ``island'', which stands out because it is directly related to the \textit{travel} genre and may be an OOD error. Under \textit{Top 5 High-Level Features}, which shows the pre-defined features associated with the highest error rates, she see that the model has the greatest error rate when \textit{overlap=Low}, which indicates when the word overlap between a \textit{premise} and a \textit{hypothesis} is low. 

Alice then focuses on a specific subpopulation to validate the actual error causes related to the presence of the \textit{island} token. From the \textit{rule discovery view} (Fig.~\ref{fig:mnli}b), she selects the rule of ``contains \textit{island}'' in order to view the examples in that subpopulation and better understand whether the errors may be caused by OOD data (\textbf{G3})
Alice first takes a look at the high-level statistics to gain an overview of the ``island'' subpopulation. In Fig.~\ref{fig:mnli}c in the \textit{Subpopulation Statistics} tab she sees that the size of the subpopulation in the training set (\textit{government} genre) is extremely small, with just 15 examples containing the word ``island''. Also, most of the errors appear when the model predicts \textit{neutral}, possibly because the model has low confidence about the relationship between hypothesis and premise in this subpopulation.


Next, she still wants to explore what kind of relationship the model needs to learn to improve the robustness. These relationships may not be learned well (\textbf{G2}) in the training or may be related to unseen data (\textbf{G3}). So she refers to the \textit{document detail view}. After reading the actual sentences that contain ``island'', she realizes that the errors may be caused by a combination of factors. The typical examples are listed in Fig.~\ref{fig:mnli}d1-d3. In the first case (Fig.~\ref{fig:mnli}d1), the model may not link ``botanical estate'' with ``many flowers''; and in the second case (Fig.~\ref{fig:mnli}d2), the model may not know that ``Milos'' is not in ``Africa''. The errors in both cases may be due to OOD issues. Specifically, the model may not know botanical concepts and geographical relationships. However, in the third case (Fig.~\ref{fig:mnli}d3), the error may occur due to the model not linking ``financial center'' with ``banks'' and ``financial institutions'' which is supposed to be covered in the government related materials. To further confirm this, Alice creates a rule of ``contain \textit{financial}'' to test this finding (\textbf{G4}) and finds that ``financial'' appears more than 1000 times in the training data (Fig.~\ref{fig:mnli}e) which is not an OOD issue. Alice continued to inspected several rules and documents. She then made a summary on the robustness test. In short, Alice finds that there are some OOD issues for the model on the travel-related text and there are still some errors caused by knowledge that the model did not learn well during the training. She decides to fine-tune her model with some geographical knowledge because this is potentially important for the government documents. Meanwhile, she decides to further inspect more cases such as finance-related words and phrases in the text from government genre to improve the model performance.

\subsection{Usage Scenario: Sentiment Analysis on Twitter}
In this hypothetical scenario, we illustrate the case when a product manager, who does not have a technical background, needs to understand when the model makes mistakes before the model is actually deployed.

A product manager, Bob, wants to apply an open-sourced model \texttt{twitter-roberta-base-sentiment}~\cite{barbieri2020tweeteval} for sentiment analysis on twitter data~\cite{barbieri2020tweeteval}. Before the actual model deployment in his product, he wants to understand where the model makes errors and wants to test a few sensitive cases he used to have trouble with. He first looks at the automatically extracted rules (Fig.~\ref{fig:ui}\textcircled{2}) to gain an overview of where the model makes more mistakes (\textbf{G1}). 

\begin{figure}
    \centering
    \includegraphics[width=0.27\textwidth]{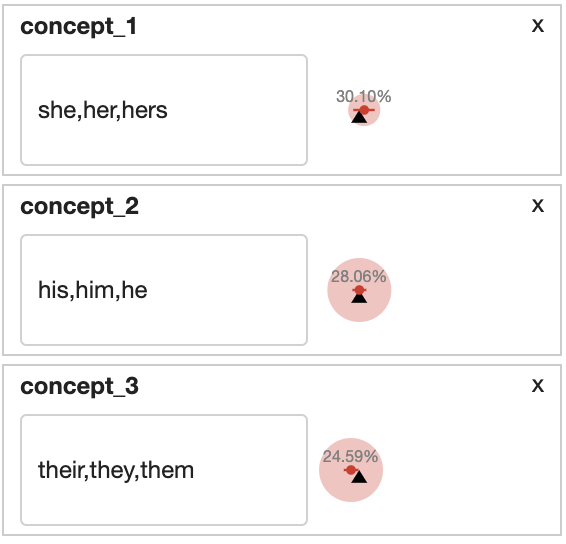}
    \caption{Three concepts of gender-related pronouns in twitter data. The subpopulations that contain such concepts are different from each other in terms of subpopulation size and error rate.}
    \Description[The view of three concepts]{The view of three concepts that refer to three different gender-related pronouns.}
    \label{fig:concept}
\end{figure}

To reason about the errors (\textbf{G2,G3}), he starts inspection of specific subpopulations. He sets the number of conditions to 1 to filter simple rules that lead to high error rate. Then he inspects the rule for phrases containing ``isis''. Under the \textit{subpopulation stat.} tab, he notices that the distribution of labels changes between the training and testing set in terms of the number of tweets containing ``isis''. The primary reason for this data shift is that the training set is based on tweets from 2013 to 2016, while the test set is from 2017 and there were not many cases containing ``isis'' in the training set as shown in Fig.~\ref{fig:ui}a). However, although only a few cases appeared in the training set, the model still learns a strong correlation between ``isis'' and a negative sentiment as shown in the aggregated bar chart of SHAP values (Fig.~\ref{fig:ui}b). He finds that the token ``isis'' increases the probability of predicting negative sentiment and decreases the probability of positive sentiment. This is a spurious correlation because after reading the tweets, he notices several cases that relay news stories about ISIS, which are neutral. Next, Bob inspects the errors in the documents containing ``kim\_jong\_un'' (R11 in Fig.~\ref{fig:ui}\textcircled{2}). He then finds that there is no such name in the training set. Because the model did not see such data before, it may not be able to make a good prediction of sentiment and thus produces neutral predictions. After inspecting more cases, Bob lists a few takeaways of using this model. For example, there are a few cases that may need to involve human input, and some tweets may contain important tokens, e.g. entities, that do not appear in the training set.

In the \textit{model performance view}, Bob notices that the model has low performance when a tweet contains a high percentage of pronouns. He then wants to test a few concepts of pronouns for different genders to check whether the model has different performance based on gender (\textbf{G4}). He thus creates a concept list as shown in Fig.~\ref{fig:concept} and finds that the subpopulation that contains the three concepts are different in size. Although the female pronoun concept (concept\_1) has a higher error rate, there is no significant difference between the error rates for this subpopulation and the male pronoun (concept\_2) and a non-binary or plural pronoun (concept\_3) as can be seen by the overlap between the 95\% confidence intervals. Based on these initial findings, further analysis may be required to more rigorously assess gender bias.


\subsection{Evaluation via Expert Interviews}
To evaluate how well {\app} can support error analysis in practice and how people use {\app}, we conduct in-depth interviews with three domain experts (E1, E2, E3) from a commercial software company. They are all research engineers that specialize in NLP (2 female, 1 male). 

\textbf{Procedure.} We conducted the interview in the form of a case study with a real-word Twitter dataset for sentiment analysis~\cite{barbieri2020tweeteval} and an open-sourced sentiment classification model provided by the authors of the dataset. Interviews were conducted remotely and lasted $45$ minutes each.
We began each interview with an introduction, during which we clarified the goal of {\app} and provided a tutorial regarding the usage of the tool. Then we asked the experts to conduct an error analysis task on the Twitter data to determine where and how the model makes mistakes. In the process, we instructed the experts to follow a ``think-aloud'' protocol~\cite{fonteyn1993description} in which they reason out loud and explicitly mention what questions they were trying to answer during the exploration and what insights they gleaned. In the final phase, we conducted a semi-structured interview which incorporated several questions about the overall usefulness, and general pros and cons of \app.

\textbf{Exploration of tool.}
In general, all the experts were able to finish the analytical task with {\app} and make use of all the functions in {\app}. All the experts went through the three stages of \textit{learning}, \textit{validating} and \textit{hypothesis testing}. They spent most time in the \textit{document detail view} to read the actual documents and reason about the SHAP values. However, we still observed some different patterns of using the tool. E1 and E3 spent more time on the concept creation view, while E2 focused more on the statistics view.




\textbf{Feature comparison.}
When asked about the most useful features of the tool, E2 and E3 listed the rule discovery view. E2 liked that the discovered rules provided a guide for further explorations. During the error analysis phase of the interview, the rule discovery view also inspired two of the experts when constructing concepts, as they chose combinations of words they had previously seen among the discovered rules.
E1 and E2 also described the document detail view as a favorite feature. E1 in fact stated that they would spend the majority of their time exploring individual examples and liked having the ability to search for examples of a particular type for model debugging. E1 and E2 even found that there may be problems with the original labeling in some subpopulations. 
E1 and E3 liked the document projection view, although were not as certain how they would directly apply it. E1 mentioned that this view would be most useful if the projection provided a separation between examples with and without errors.

\textbf{Concept creation.}
When queried about concepts that they would find most useful for hypothesis testing, all three experts mentioned concepts related to model bias, for example race or gender. E3 was broadly interested in different types of entities, such as places and person names. E2 also expressed interest in collections of hashtags describing particular events or topics.


\textbf{Suggestions for improvement.} 
We also prompted the experts for suggestions on how to improve the tool. E1 suggested supporting regular expression for concept construction. E2 and E3 mentioned that other users might not be familiar with some terms such as SHAP value, subpopulation, and concept, and they suggested including more guidance from the tool, e.g. tooltips, to explain these terms. E2 found that the interface tried to leverage a lot of different statistics and suggested grouping similar things. E2 and E3 also suggested simplifying the visualization of concept summary.  E2 also mentioned that it was challenging to interpret some of the extracted rules, for example, rules containing prepositions\footnote{We note that some prespositions such as ``from'' and ``after'' are related to location and time logic which can be useful in some cases, while others may represent noise in the data that should be filtered out.}.

%% file: sections/07-limit+conclude.tex
\section{Discussion}
As a pipeline for error analysis of NLP models, we hope that {\app} will have a positive impact on the development and deployment of NLP models or even other types of models. We discuss the impact with respect to the following aspects:


\textbf{Enabling semantic error analysis without coding.}
Enabling people to analyze model behaviors, especially erroneous behaviors increases the transparency and fairness of the whole machine learning pipeline. The user interface of {\app} enables all the stakeholders, who even do not have a technical background, to understand the model mistakes without any coding. 
Although we present the evaluation of two usage scenarios of model development and deployment, we expect that {\app} can be used even by the end-users of the actual AI techniques with proper guidance and explanations to increase their trust of the model decisions, be aware of model uncertainty, and understand when user input is necessary. 

\textbf{Importance of the human-in-the-loop pipeline with intelligent UI.}
We also confirm the importance of involving humans in the loop with the assistance of an intelligent UI for error analysis through the development of this work. Although the automatically extracted rules provide a description of error-prone subpopulations, they do not reveal the underlying reason for the errors. For example, the errors related to some person names may be caused by an OOD issue or incorrect labeling. These actual causes of errors need further analysis and validation from a human user. 
Regardless, these automatically extracted error explanations still provide value during the error analysis, because they guide the users to a subpopulation that should be investigated and inspire the users to reason about the errors and create their own rules. We expect to see more work that integrates human and machine intelligence in error analysis.

\textbf{Understanding model bias.}
The interviewed experts showed an interest in testing and comparing the model performance in terms of different tokens or concepts. They mentioned concepts related to gender, race, and geographical locations, for example.
Although a more systematic exploration and evaluation of such impact is needed, it is good to see that people think of such comparison during the usage of {\app}.  We expect {\app} to encourage model developers to test and improve the model performance for a more diverse group of people and data.

\textbf{Extension to error analysis of other models.}
Although this work focuses on the error analysis of NLP models, the presented error analysis pipeline and system architecture can be widely applicable. The high-level synthesis of building an error analysis system can be summarized as follows: (1) the pipeline as shown in Fig.~\ref{fig:pipeline} illustrates the key points of \textit{learning}, \textit{validating}, and \textit{hypothesis testing} in a procedure of error analysis that involves humans in the loop, which can be extended to other model debugging and model diagnosis problems; (2) the rule presentation principles for errors, such as \textit{``limit the number of conditions and cardinality of features''} and \textit{``test significance''} are also applicable to other types of model and data, while \textit{``avoid negation''} may be unnecessary for other data because the negative values can still be meaningful in many cases; (3) the system architecture we described in Fig.~\ref{fig:data_process} provides a manageable method that integrates model, analysis process, and pre-mined results with an interactive user interface, which can be applied to other analysis problems related to machine learning models.

\section{Limitations}
This work is the first step in our goal to provide a full user-centered error analysis tool. The system currently has certain limitations. The first limitation is the understanding of complex semantics and context of a document. In {\app}, we use token-level features to discover semantically-grounded subpopulations that contain errors. However, token-level features may ignore the context of the text. For example,
``This is not her best work.'' and ``This is her best work, not to be missed.'' share similar vocabulary (tokens) but have quite different semantic meanings. To better capture the semantics in the documents, we include concepts and high-level features (e.g., number of adjectives) in the system, which supports more flexible subpopulation discovery and construction. Although these features are complementary to tokens, the context of a document still may not be well depicted. More research is needed to explore interpretable features and representations that may assist users in understanding more complex semantics in their full context.

Further, {\app} only supports error analysis of classification tasks that require textual information, including sentiment analysis, NLI, text classification, and yes/no question answering. The tasks such as VQA which involves image information, and translation which related to text generation are not supported at present. 
Finally, by only interviewing three domain experts, we may be overgeneralizing our results.

\section{Conclusion and Future Work}
In this paper, we present an interactive pipeline for users to discover and analyze the errors made by an NLP model. To extract and present rules that describe error-prone subpopulations with improved interpretability, we developed three different types of features for rules and followed four principles of rule generation and presentation. As a result of this, we designed and implemented {\app}, a graphical interface for semantic error analysis that allows users to \textit{learn} and \textit{validate} the causes of errors, as well as \textit{test hypothesis} on model performance. We demonstrated the usefulness of {\app} through two usage scenarios with two NLP tasks: NLI and semantic analysis, and in-depth interviews with domain experts. 
In the future, we plan to apply {\app} to more NLP tasks, and develop more intelligent assistance in concept creation and evaluation. Moreover, we intend to conduct a formal user study to demonstrate the usability of different views in our system and compare our approach with existing error analysis tools.

%% file: main.bbl

\begin{thebibliography}{33}


\ifx \showCODEN    \undefined \def \showCODEN     #1{\unskip}     \fi
\ifx \showDOI      \undefined \def \showDOI       #1{#1}\fi
\ifx \showISBNx    \undefined \def \showISBNx     #1{\unskip}     \fi
\ifx \showISBNxiii \undefined \def \showISBNxiii  #1{\unskip}     \fi
\ifx \showISSN     \undefined \def \showISSN      #1{\unskip}     \fi
\ifx \showLCCN     \undefined \def \showLCCN      #1{\unskip}     \fi
\ifx \shownote     \undefined \def \shownote      #1{#1}          \fi
\ifx \showarticletitle \undefined \def \showarticletitle #1{#1}   \fi
\ifx \showURL      \undefined \def \showURL       {\relax}        \fi
\providecommand\bibfield[2]{#2}
\providecommand\bibinfo[2]{#2}
\providecommand\natexlab[1]{#1}
\providecommand\showeprint[2][]{arXiv:#2}

\bibitem[\protect\citeauthoryear{Alsallakh, Hanbury, Hauser, Miksch, and
  Rauber}{Alsallakh et~al\mbox{.}}{2014}]%
        {alsallakh2014visual}
\bibfield{author}{\bibinfo{person}{Bilal Alsallakh}, \bibinfo{person}{Allan
  Hanbury}, \bibinfo{person}{Helwig Hauser}, \bibinfo{person}{Silvia Miksch},
  {and} \bibinfo{person}{Andreas Rauber}.} \bibinfo{year}{2014}\natexlab{}.
\newblock \showarticletitle{Visual methods for analyzing probabilistic
  classification data}.
\newblock \bibinfo{journal}{\emph{IEEE transactions on visualization and
  computer graphics}} \bibinfo{volume}{20}, \bibinfo{number}{12}
  (\bibinfo{year}{2014}), \bibinfo{pages}{1703--1712}.
\newblock


\bibitem[\protect\citeauthoryear{Arendt, Huang, Shrestha, Ayton, Glenski, and
  Volkova}{Arendt et~al\mbox{.}}{2020}]%
        {arendt2020crosscheck}
\bibfield{author}{\bibinfo{person}{Dustin Arendt}, \bibinfo{person}{Zhuanyi
  Huang}, \bibinfo{person}{Prasha Shrestha}, \bibinfo{person}{Ellyn Ayton},
  \bibinfo{person}{Maria Glenski}, {and} \bibinfo{person}{Svitlana Volkova}.}
  \bibinfo{year}{2020}\natexlab{}.
\newblock \showarticletitle{Crosscheck: Rapid, reproducible, and interpretable
  model evaluation}.
\newblock \bibinfo{journal}{\emph{arXiv preprint arXiv:2004.07993}}
  (\bibinfo{year}{2020}).
\newblock


\bibitem[\protect\citeauthoryear{Barbieri, Camacho-Collados, Espinosa-Anke, and
  Neves}{Barbieri et~al\mbox{.}}{2020}]%
        {barbieri2020tweeteval}
\bibfield{author}{\bibinfo{person}{Francesco Barbieri}, \bibinfo{person}{Jose
  Camacho-Collados}, \bibinfo{person}{Luis Espinosa-Anke}, {and}
  \bibinfo{person}{Leonardo Neves}.} \bibinfo{year}{2020}\natexlab{}.
\newblock \showarticletitle{{TweetEval:Unified Benchmark and Comparative
  Evaluation for Tweet Classification}}. In
  \bibinfo{booktitle}{\emph{Proceedings of Findings of EMNLP}}.
\newblock


\bibitem[\protect\citeauthoryear{Bilal, Jourabloo, Ye, Liu, and Ren}{Bilal
  et~al\mbox{.}}{2017}]%
        {bilal2017convolutional}
\bibfield{author}{\bibinfo{person}{Alsallakh Bilal}, \bibinfo{person}{Amin
  Jourabloo}, \bibinfo{person}{Mao Ye}, \bibinfo{person}{Xiaoming Liu}, {and}
  \bibinfo{person}{Liu Ren}.} \bibinfo{year}{2017}\natexlab{}.
\newblock \showarticletitle{Do convolutional neural networks learn class
  hierarchy?}
\newblock \bibinfo{journal}{\emph{IEEE transactions on visualization and
  computer graphics}} \bibinfo{volume}{24}, \bibinfo{number}{1}
  (\bibinfo{year}{2017}), \bibinfo{pages}{152--162}.
\newblock


\bibitem[\protect\citeauthoryear{Cabrera, Epperson, Hohman, Kahng, Morgenstern,
  and Chau}{Cabrera et~al\mbox{.}}{2019}]%
        {cabrera2019fairvis}
\bibfield{author}{\bibinfo{person}{{\'A}ngel~Alexander Cabrera},
  \bibinfo{person}{Will Epperson}, \bibinfo{person}{Fred Hohman},
  \bibinfo{person}{Minsuk Kahng}, \bibinfo{person}{Jamie Morgenstern}, {and}
  \bibinfo{person}{Duen~Horng Chau}.} \bibinfo{year}{2019}\natexlab{}.
\newblock \showarticletitle{FairVis: Visual analytics for discovering
  intersectional bias in machine learning}. In \bibinfo{booktitle}{\emph{2019
  IEEE Conference on Visual Analytics Science and Technology (VAST)}}. IEEE,
  \bibinfo{pages}{46--56}.
\newblock


\bibitem[\protect\citeauthoryear{Chen, Gu, Ji, Lou, Sun, Li, Gao, and
  Huang}{Chen et~al\mbox{.}}{2019}]%
        {chen2019clinical}
\bibfield{author}{\bibinfo{person}{Long Chen}, \bibinfo{person}{Yu Gu},
  \bibinfo{person}{Xin Ji}, \bibinfo{person}{Chao Lou},
  \bibinfo{person}{Zhiyong Sun}, \bibinfo{person}{Haodan Li},
  \bibinfo{person}{Yuan Gao}, {and} \bibinfo{person}{Yang Huang}.}
  \bibinfo{year}{2019}\natexlab{}.
\newblock \showarticletitle{Clinical trial cohort selection based on
  multi-level rule-based natural language processing system}.
\newblock \bibinfo{journal}{\emph{Journal of the American Medical Informatics
  Association}} \bibinfo{volume}{26}, \bibinfo{number}{11}
  (\bibinfo{year}{2019}), \bibinfo{pages}{1218--1226}.
\newblock


\bibitem[\protect\citeauthoryear{Chung, Kraska, Polyzotis, Tae, and
  Whang}{Chung et~al\mbox{.}}{2019}]%
        {chung2019slice}
\bibfield{author}{\bibinfo{person}{Yeounoh Chung}, \bibinfo{person}{Tim
  Kraska}, \bibinfo{person}{Neoklis Polyzotis}, \bibinfo{person}{Ki~Hyun Tae},
  {and} \bibinfo{person}{Steven~Euijong Whang}.}
  \bibinfo{year}{2019}\natexlab{}.
\newblock \showarticletitle{Slice finder: Automated data slicing for model
  validation}. In \bibinfo{booktitle}{\emph{2019 IEEE 35th International
  Conference on Data Engineering (ICDE)}}. IEEE, \bibinfo{pages}{1550--1553}.
\newblock


\bibitem[\protect\citeauthoryear{DiCiccio and Efron}{DiCiccio and
  Efron}{1996}]%
        {diciccio1996bootstrap}
\bibfield{author}{\bibinfo{person}{Thomas~J DiCiccio} {and}
  \bibinfo{person}{Bradley Efron}.} \bibinfo{year}{1996}\natexlab{}.
\newblock \showarticletitle{Bootstrap confidence intervals}.
\newblock \bibinfo{journal}{\emph{Statistical science}} \bibinfo{volume}{11},
  \bibinfo{number}{3} (\bibinfo{year}{1996}), \bibinfo{pages}{189--228}.
\newblock


\bibitem[\protect\citeauthoryear{Fonteyn, Kuipers, and Grobe}{Fonteyn
  et~al\mbox{.}}{1993}]%
        {fonteyn1993description}
\bibfield{author}{\bibinfo{person}{Marsha~E Fonteyn}, \bibinfo{person}{Benjamin
  Kuipers}, {and} \bibinfo{person}{Susan~J Grobe}.}
  \bibinfo{year}{1993}\natexlab{}.
\newblock \showarticletitle{A description of think aloud method and protocol
  analysis}.
\newblock \bibinfo{journal}{\emph{Qualitative health research}}
  \bibinfo{volume}{3}, \bibinfo{number}{4} (\bibinfo{year}{1993}),
  \bibinfo{pages}{430--441}.
\newblock


\bibitem[\protect\citeauthoryear{Garrido-Mu{\~n}oz, Montejo-R{\'a}ez,
  Mart{\'\i}nez-Santiago, and Ure{\~n}a-L{\'o}pez}{Garrido-Mu{\~n}oz
  et~al\mbox{.}}{2021}]%
        {garrido2021survey}
\bibfield{author}{\bibinfo{person}{Ismael Garrido-Mu{\~n}oz},
  \bibinfo{person}{Arturo Montejo-R{\'a}ez}, \bibinfo{person}{Fernando
  Mart{\'\i}nez-Santiago}, {and} \bibinfo{person}{L~Alfonso
  Ure{\~n}a-L{\'o}pez}.} \bibinfo{year}{2021}\natexlab{}.
\newblock \showarticletitle{A Survey on Bias in Deep NLP}.
\newblock \bibinfo{journal}{\emph{Applied Sciences}} \bibinfo{volume}{11},
  \bibinfo{number}{7} (\bibinfo{year}{2021}), \bibinfo{pages}{3184}.
\newblock


\bibitem[\protect\citeauthoryear{Goel, Rajani, Vig, Tan, Wu, Zheng, Xiong,
  Bansal, and R{\'e}}{Goel et~al\mbox{.}}{2021}]%
        {goel2021robustness}
\bibfield{author}{\bibinfo{person}{Karan Goel}, \bibinfo{person}{Nazneen
  Rajani}, \bibinfo{person}{Jesse Vig}, \bibinfo{person}{Samson Tan},
  \bibinfo{person}{Jason Wu}, \bibinfo{person}{Stephan Zheng},
  \bibinfo{person}{Caiming Xiong}, \bibinfo{person}{Mohit Bansal}, {and}
  \bibinfo{person}{Christopher R{\'e}}.} \bibinfo{year}{2021}\natexlab{}.
\newblock \showarticletitle{Robustness gym: Unifying the nlp evaluation
  landscape}.
\newblock \bibinfo{journal}{\emph{arXiv preprint arXiv:2101.04840}}
  (\bibinfo{year}{2021}).
\newblock


\bibitem[\protect\citeauthoryear{Gururangan, Swayamdipta, Levy, Schwartz,
  Bowman, and Smith}{Gururangan et~al\mbox{.}}{2018}]%
        {gururangan2018annotation}
\bibfield{author}{\bibinfo{person}{Suchin Gururangan}, \bibinfo{person}{Swabha
  Swayamdipta}, \bibinfo{person}{Omer Levy}, \bibinfo{person}{Roy Schwartz},
  \bibinfo{person}{Samuel~R Bowman}, {and} \bibinfo{person}{Noah~A Smith}.}
  \bibinfo{year}{2018}\natexlab{}.
\newblock \showarticletitle{Annotation artifacts in natural language inference
  data}.
\newblock \bibinfo{journal}{\emph{arXiv preprint arXiv:1803.02324}}
  (\bibinfo{year}{2018}).
\newblock


\bibitem[\protect\citeauthoryear{Hutto and Gilbert}{Hutto and Gilbert}{2014}]%
        {hutto2014vader}
\bibfield{author}{\bibinfo{person}{Clayton Hutto} {and} \bibinfo{person}{Eric
  Gilbert}.} \bibinfo{year}{2014}\natexlab{}.
\newblock \showarticletitle{Vader: A parsimonious rule-based model for
  sentiment analysis of social media text}. In
  \bibinfo{booktitle}{\emph{Proceedings of the International AAAI Conference on
  Web and Social Media}}, Vol.~\bibinfo{volume}{8}.
\newblock


\bibitem[\protect\citeauthoryear{Joshi, Dai, Karimi, Sparks, Paris, and
  MacIntyre}{Joshi et~al\mbox{.}}{2018}]%
        {joshi2018shot}
\bibfield{author}{\bibinfo{person}{Aditya Joshi}, \bibinfo{person}{Xiang Dai},
  \bibinfo{person}{Sarvnaz Karimi}, \bibinfo{person}{Ross Sparks},
  \bibinfo{person}{Cecile Paris}, {and} \bibinfo{person}{C~Raina MacIntyre}.}
  \bibinfo{year}{2018}\natexlab{}.
\newblock \showarticletitle{Shot or not: Comparison of NLP approaches for
  vaccination behaviour detection}. In \bibinfo{booktitle}{\emph{Proceedings of
  the 2018 EMNLP Workshop SMM4H: The 3rd Social Media Mining for Health
  Applications Workshop \& Shared Task}}. \bibinfo{pages}{43--47}.
\newblock


\bibitem[\protect\citeauthoryear{Kahng, Andrews, Kalro, and Chau}{Kahng
  et~al\mbox{.}}{2017}]%
        {kahng2017cti}
\bibfield{author}{\bibinfo{person}{Minsuk Kahng}, \bibinfo{person}{Pierre~Y
  Andrews}, \bibinfo{person}{Aditya Kalro}, {and} \bibinfo{person}{Duen~Horng
  Chau}.} \bibinfo{year}{2017}\natexlab{}.
\newblock \showarticletitle{A cti v is: Visual exploration of industry-scale
  deep neural network models}.
\newblock \bibinfo{journal}{\emph{IEEE transactions on visualization and
  computer graphics}} \bibinfo{volume}{24}, \bibinfo{number}{1}
  (\bibinfo{year}{2017}), \bibinfo{pages}{88--97}.
\newblock


\bibitem[\protect\citeauthoryear{Li, Ji, Du, Li, and Wang}{Li
  et~al\mbox{.}}{2018}]%
        {li2018textbugger}
\bibfield{author}{\bibinfo{person}{Jinfeng Li}, \bibinfo{person}{Shouling Ji},
  \bibinfo{person}{Tianyu Du}, \bibinfo{person}{Bo Li}, {and}
  \bibinfo{person}{Ting Wang}.} \bibinfo{year}{2018}\natexlab{}.
\newblock \showarticletitle{Textbugger: Generating adversarial text against
  real-world applications}.
\newblock \bibinfo{journal}{\emph{arXiv preprint arXiv:1812.05271}}
  (\bibinfo{year}{2018}).
\newblock


\bibitem[\protect\citeauthoryear{Liu, Shen, Duh, and Gao}{Liu
  et~al\mbox{.}}{2017}]%
        {liu2017stochastic}
\bibfield{author}{\bibinfo{person}{Xiaodong Liu}, \bibinfo{person}{Yelong
  Shen}, \bibinfo{person}{Kevin Duh}, {and} \bibinfo{person}{Jianfeng Gao}.}
  \bibinfo{year}{2017}\natexlab{}.
\newblock \showarticletitle{Stochastic answer networks for machine reading
  comprehension}.
\newblock \bibinfo{journal}{\emph{arXiv preprint arXiv:1712.03556}}
  (\bibinfo{year}{2017}).
\newblock


\bibitem[\protect\citeauthoryear{Lundberg and Lee}{Lundberg and Lee}{2017}]%
        {lundberg2017unified}
\bibfield{author}{\bibinfo{person}{Scott~M Lundberg} {and}
  \bibinfo{person}{Su-In Lee}.} \bibinfo{year}{2017}\natexlab{}.
\newblock \showarticletitle{A unified approach to interpreting model
  predictions}. In \bibinfo{booktitle}{\emph{Proceedings of the 31st
  international conference on neural information processing systems}}.
  \bibinfo{pages}{4768--4777}.
\newblock


\bibitem[\protect\citeauthoryear{McCoy, Pavlick, and Linzen}{McCoy
  et~al\mbox{.}}{2019}]%
        {mccoy2019right}
\bibfield{author}{\bibinfo{person}{R~Thomas McCoy}, \bibinfo{person}{Ellie
  Pavlick}, {and} \bibinfo{person}{Tal Linzen}.}
  \bibinfo{year}{2019}\natexlab{}.
\newblock \showarticletitle{Right for the wrong reasons: Diagnosing syntactic
  heuristics in natural language inference}.
\newblock \bibinfo{journal}{\emph{arXiv preprint arXiv:1902.01007}}
  (\bibinfo{year}{2019}).
\newblock


\bibitem[\protect\citeauthoryear{Microsoft}{Microsoft}{2021}]%
        {respobibleai}
\bibfield{author}{\bibinfo{person}{Microsoft}.}
  \bibinfo{year}{2021}\natexlab{}.
\newblock \bibinfo{title}{microsoft/responsible-ai-widgets}.
\newblock
  \bibinfo{howpublished}{\url{https://github.com/microsoft/responsible-ai-widgets}}.
\newblock
\newblock
\shownote{Accessed: 2021-10-08.}


\bibitem[\protect\citeauthoryear{Reimers and Gurevych}{Reimers and
  Gurevych}{2019}]%
        {reimers2019sentence}
\bibfield{author}{\bibinfo{person}{Nils Reimers} {and} \bibinfo{person}{Iryna
  Gurevych}.} \bibinfo{year}{2019}\natexlab{}.
\newblock \showarticletitle{Sentence-bert: Sentence embeddings using siamese
  bert-networks}.
\newblock \bibinfo{journal}{\emph{arXiv preprint arXiv:1908.10084}}
  (\bibinfo{year}{2019}).
\newblock


\bibitem[\protect\citeauthoryear{Ribeiro, Singh, and Guestrin}{Ribeiro
  et~al\mbox{.}}{2016}]%
        {ribeiro2016should}
\bibfield{author}{\bibinfo{person}{Marco~Tulio Ribeiro},
  \bibinfo{person}{Sameer Singh}, {and} \bibinfo{person}{Carlos Guestrin}.}
  \bibinfo{year}{2016}\natexlab{}.
\newblock \showarticletitle{" Why should i trust you?" Explaining the
  predictions of any classifier}. In \bibinfo{booktitle}{\emph{Proceedings of
  the 22nd ACM SIGKDD international conference on knowledge discovery and data
  mining}}. \bibinfo{pages}{1135--1144}.
\newblock


\bibitem[\protect\citeauthoryear{Ribeiro, Wu, Guestrin, and Singh}{Ribeiro
  et~al\mbox{.}}{2020}]%
        {ribeiro2020beyond}
\bibfield{author}{\bibinfo{person}{Marco~Tulio Ribeiro},
  \bibinfo{person}{Tongshuang Wu}, \bibinfo{person}{Carlos Guestrin}, {and}
  \bibinfo{person}{Sameer Singh}.} \bibinfo{year}{2020}\natexlab{}.
\newblock \showarticletitle{Beyond accuracy: Behavioral testing of NLP models
  with CheckList}.
\newblock \bibinfo{journal}{\emph{arXiv preprint arXiv:2005.04118}}
  (\bibinfo{year}{2020}).
\newblock


\bibitem[\protect\citeauthoryear{Sanh, Debut, Chaumond, and Wolf}{Sanh
  et~al\mbox{.}}{2019}]%
        {sanh2019distilbert}
\bibfield{author}{\bibinfo{person}{Victor Sanh}, \bibinfo{person}{Lysandre
  Debut}, \bibinfo{person}{Julien Chaumond}, {and} \bibinfo{person}{Thomas
  Wolf}.} \bibinfo{year}{2019}\natexlab{}.
\newblock \showarticletitle{DistilBERT, a distilled version of BERT: smaller,
  faster, cheaper and lighter}.
\newblock \bibinfo{journal}{\emph{arXiv preprint arXiv:1910.01108}}
  (\bibinfo{year}{2019}).
\newblock


\bibitem[\protect\citeauthoryear{Shrikumar, Greenside, Shcherbina, and
  Kundaje}{Shrikumar et~al\mbox{.}}{2016}]%
        {shrikumar2016not}
\bibfield{author}{\bibinfo{person}{Avanti Shrikumar}, \bibinfo{person}{Peyton
  Greenside}, \bibinfo{person}{Anna Shcherbina}, {and} \bibinfo{person}{Anshul
  Kundaje}.} \bibinfo{year}{2016}\natexlab{}.
\newblock \showarticletitle{Not just a black box: Learning important features
  through propagating activation differences}.
\newblock \bibinfo{journal}{\emph{arXiv preprint arXiv:1605.01713}}
  (\bibinfo{year}{2016}).
\newblock


\bibitem[\protect\citeauthoryear{Tan and Celis}{Tan and Celis}{2019}]%
        {tan2019assessing}
\bibfield{author}{\bibinfo{person}{Yi~Chern Tan} {and} \bibinfo{person}{L~Elisa
  Celis}.} \bibinfo{year}{2019}\natexlab{}.
\newblock \showarticletitle{Assessing social and intersectional biases in
  contextualized word representations}.
\newblock \bibinfo{journal}{\emph{arXiv preprint arXiv:1911.01485}}
  (\bibinfo{year}{2019}).
\newblock


\bibitem[\protect\citeauthoryear{Tenney, Wexler, Bastings, Bolukbasi, Coenen,
  Gehrmann, Jiang, Pushkarna, Radebaugh, Reif, et~al\mbox{.}}{Tenney
  et~al\mbox{.}}{2020}]%
        {tenney2020language}
\bibfield{author}{\bibinfo{person}{Ian Tenney}, \bibinfo{person}{James Wexler},
  \bibinfo{person}{Jasmijn Bastings}, \bibinfo{person}{Tolga Bolukbasi},
  \bibinfo{person}{Andy Coenen}, \bibinfo{person}{Sebastian Gehrmann},
  \bibinfo{person}{Ellen Jiang}, \bibinfo{person}{Mahima Pushkarna},
  \bibinfo{person}{Carey Radebaugh}, \bibinfo{person}{Emily Reif},
  {et~al\mbox{.}}} \bibinfo{year}{2020}\natexlab{}.
\newblock \showarticletitle{The language interpretability tool: Extensible,
  interactive visualizations and analysis for NLP models}.
\newblock \bibinfo{journal}{\emph{arXiv preprint arXiv:2008.05122}}
  (\bibinfo{year}{2020}).
\newblock


\bibitem[\protect\citeauthoryear{Van~der Maaten and Hinton}{Van~der Maaten and
  Hinton}{2008}]%
        {van2008visualizing}
\bibfield{author}{\bibinfo{person}{Laurens Van~der Maaten} {and}
  \bibinfo{person}{Geoffrey Hinton}.} \bibinfo{year}{2008}\natexlab{}.
\newblock \showarticletitle{Visualizing data using t-SNE.}
\newblock \bibinfo{journal}{\emph{Journal of machine learning research}}
  \bibinfo{volume}{9}, \bibinfo{number}{11} (\bibinfo{year}{2008}).
\newblock


\bibitem[\protect\citeauthoryear{Vig, Gehrmann, Belinkov, Qian, Nevo, Sakenis,
  Huang, Singer, and Shieber}{Vig et~al\mbox{.}}{2020}]%
        {vig2020causal}
\bibfield{author}{\bibinfo{person}{Jesse Vig}, \bibinfo{person}{Sebastian
  Gehrmann}, \bibinfo{person}{Yonatan Belinkov}, \bibinfo{person}{Sharon Qian},
  \bibinfo{person}{Daniel Nevo}, \bibinfo{person}{Simas Sakenis},
  \bibinfo{person}{Jason Huang}, \bibinfo{person}{Yaron Singer}, {and}
  \bibinfo{person}{Stuart Shieber}.} \bibinfo{year}{2020}\natexlab{}.
\newblock \showarticletitle{Causal mediation analysis for interpreting neural
  nlp: The case of gender bias}.
\newblock \bibinfo{journal}{\emph{arXiv preprint arXiv:2004.12265}}
  (\bibinfo{year}{2020}).
\newblock


\bibitem[\protect\citeauthoryear{Wang, Gou, Zhang, Yang, and Shen}{Wang
  et~al\mbox{.}}{2019}]%
        {wang2019deepvid}
\bibfield{author}{\bibinfo{person}{Junpeng Wang}, \bibinfo{person}{Liang Gou},
  \bibinfo{person}{Wei Zhang}, \bibinfo{person}{Hao Yang}, {and}
  \bibinfo{person}{Han-Wei Shen}.} \bibinfo{year}{2019}\natexlab{}.
\newblock \showarticletitle{Deepvid: Deep visual interpretation and diagnosis
  for image classifiers via knowledge distillation}.
\newblock \bibinfo{journal}{\emph{IEEE transactions on visualization and
  computer graphics}} \bibinfo{volume}{25}, \bibinfo{number}{6}
  (\bibinfo{year}{2019}), \bibinfo{pages}{2168--2180}.
\newblock


\bibitem[\protect\citeauthoryear{Williams, Nangia, and Bowman}{Williams
  et~al\mbox{.}}{2018}]%
        {N18-1101}
\bibfield{author}{\bibinfo{person}{Adina Williams}, \bibinfo{person}{Nikita
  Nangia}, {and} \bibinfo{person}{Samuel Bowman}.}
  \bibinfo{year}{2018}\natexlab{}.
\newblock \showarticletitle{A Broad-Coverage Challenge Corpus for Sentence
  Understanding through Inference}. In \bibinfo{booktitle}{\emph{Proceedings of
  the 2018 Conference of the North American Chapter of the Association for
  Computational Linguistics: Human Language Technologies, Volume 1 (Long
  Papers)}} (New Orleans, Louisiana). \bibinfo{publisher}{Association for
  Computational Linguistics}, \bibinfo{pages}{1112--1122}.
\newblock
\urldef\tempurl%
\url{http://aclweb.org/anthology/N18-1101}
\showURL{%
\tempurl}


\bibitem[\protect\citeauthoryear{Wu, Ribeiro, Heer, and Weld}{Wu
  et~al\mbox{.}}{2019}]%
        {wu2019errudite}
\bibfield{author}{\bibinfo{person}{Tongshuang Wu}, \bibinfo{person}{Marco~Tulio
  Ribeiro}, \bibinfo{person}{Jeffrey Heer}, {and} \bibinfo{person}{Daniel~S
  Weld}.} \bibinfo{year}{2019}\natexlab{}.
\newblock \showarticletitle{Errudite: Scalable, reproducible, and testable
  error analysis}. In \bibinfo{booktitle}{\emph{Proceedings of the 57th Annual
  Meeting of the Association for Computational Linguistics}}.
  \bibinfo{pages}{747--763}.
\newblock


\bibitem[\protect\citeauthoryear{Zhang, Wang, Molino, Li, and Ebert}{Zhang
  et~al\mbox{.}}{2018}]%
        {zhang2018manifold}
\bibfield{author}{\bibinfo{person}{Jiawei Zhang}, \bibinfo{person}{Yang Wang},
  \bibinfo{person}{Piero Molino}, \bibinfo{person}{Lezhi Li}, {and}
  \bibinfo{person}{David~S Ebert}.} \bibinfo{year}{2018}\natexlab{}.
\newblock \showarticletitle{Manifold: A model-agnostic framework for
  interpretation and diagnosis of machine learning models}.
\newblock \bibinfo{journal}{\emph{IEEE transactions on visualization and
  computer graphics}} \bibinfo{volume}{25}, \bibinfo{number}{1}
  (\bibinfo{year}{2018}), \bibinfo{pages}{364--373}.
\newblock


\end{thebibliography}
